\documentclass[twocolumn,prb,aps,showpacs]{revtex4}
\usepackage[dvips]{graphicx}
\usepackage{color}
\usepackage{hyperref}
\usepackage{amssymb}
\usepackage{amsmath}
\usepackage{overpic}

\allowdisplaybreaks

\usepackage{tikz}
\usetikzlibrary{calc,positioning}
\tikzset{%
  hiblue/.style={rectangle,rounded corners,fill=blue!55,draw,fill
opacity=1,anchor=base}
}

\tikzset{%
  hired/.style={rectangle,rounded corners,fill=red!55,draw,fill
opacity=1,anchor=base}
}

\newcommand{\hilite}[2]{%
    \tikz[baseline]{
    \node[#1] {#2};}
}

\newcommand{\anode}[2]{%
    \tikz[baseline,remember picture]{
    \node[anchor=base] (#1) {#2};}
}

\newcommand{\hilitenode}[3]{%
    \tikz[baseline,remember picture]{
    \node[#1](#2) {#3};}
}

\xdefinecolor{darkgreen}{RGB}{0, 128, 0}

\newcommand{\up}{\uparrow}
\newcommand{\dow}{\downarrow}

\begin{document}
\title{Conserved quantities of SU(2)-invariant interactions for correlated fermions and the advantages for quantum Monte Carlo simulations.}


\begin{abstract} 
In the context of realistic calculations for strongly-correlated materials with $d$- or $f$-electrons the efficient computation of multi-orbital models is of paramount importance. Here we introduce a set of invariants for the SU(2)-symmetric Kanamori Hamiltonian which allows to massively speed up the calculation of the fermionic trace in hybridization-expansion continuous-time quantum Monte Carlo algorithms.
As an application, we show that, exploiting this set of good quantum numbers, the study of the orbital-selective Mott-transition in systems with up to seven correlated orbitals becomes feasible.
\end{abstract}

\author{Nicolaus Parragh$^{(1)}$, Alessandro Toschi$^{(2)}$, Karsten Held$^{(2)}$ and Giorgio Sangiovanni$^{(1)}$}
\affiliation{
${}^1$Institut f\"ur Theoretische Physik und Astrophysik, Universit\"at W\"urzburg, Am Hubland, D-97074 W\"urzburg, Germany \\
${}^2$Institut f\"ur Festk\"orperphysik, Technische Universit\"at Wien, Vienna, Austria \\
}

\maketitle

The calculation of the electronic properties of materials with $d$- and $f$-electrons requires highly efficient numerical algorithms capable of treating systems of many interacting fermions. 
Dynamical mean field theory (DMFT) and its cluster as well as diagrammatic extensions have proven very successful in predicting one- and two-particle dynamical quantities \cite{metznerPRL62,georgesRMP68,extensions}. Moreover, the combination with density functional theory makes it possible to predict a great number of material-specific effects \cite{LDADMFT,kotliarRMP78,heldAdv56}.
These theories drastically reduce the complexity of the original lattice problem by mapping it onto an appropriate Anderson model (containing either one single impurity or a small cluster of them). This however still constitutes a highly non-trivial many-body problem, in particular when the impurity site contains more than one orbital, and therefore is the bottleneck of these methods.
Improvements in the numerical efficiency of the algorithms for solving the impurity model, like the one we discuss here, are of great importance, since they make unexplored regions of the model phase diagrams accessible and the study of new materials possible.

Recently continuous-time quantum Monte Carlo (CT-QMC) algorithms have been introduced \cite{CTQMC,wernerPRB74,haulePRB75,gullRMP83}. They represent a breakthrough in the development of efficient ``impurity solvers'' for strongly correlated electron systems. 
Already from the early stages it was clear that multi-orbital models for $d$- and $f$-electron systems with SU(2)-symmetric ``Kanamori'' kind of interactions [see Eq. (\ref{Kanamori}) below] are very well suited to be studied with CT-QMC, in particular with the hybridization-expansion (CT-HYB) \cite{wernerPRB74}.
In CT-HYB one splits the full Hamiltonian of the Anderson impurity problem into an interacting part involving the isolated impurity only ($H_{loc}$), a part for the non-interacting bath only, and a hybridization between the impurity and the bath. The bath part is analytically integrated out and the Monte Carlo simulation consists of sampling a fermionic trace in which the imaginary-time evolution between 0 and $\beta\!=\!1/T$ is governed by $H_{loc}$ and at random imaginary-time positions creation and annihilation operators for fermions on the impurity site are inserted and removed.

The standard implementation of CT-HYB is formulated in the eigenbasis of $H_{loc}$ and the trace is evaluated via a number of matrix-matrix multiplications, which is tractable for systems with up to three orbitals.
L\"auchli and Werner \cite{laeuchliPRB80} put forward a very elegant solution for simulations with more orbitals based on the Lanczos algorithm. In this so-called ``Krylov implementation'' the trace is calculated using Lanczos and fast sparse-matrix/vector operations.
Independently of the implementation used, it is clear that the more one reduces the size of the blocks of $H_{loc}$ exploiting its good quantum numbers, the faster the calculations go \cite{haulePRB75}.
Therefore, for the efficiency of the whole computational scheme, it is crucial to identify as many good quantum numbers as possible and to make sure that they can be efficiently treated by the code.
In the present paper we introduce what we call the ``{\bf PS}'' vector, a set of conserved quantities for the SU(2)-symmetric Kanamori Hamiltonian which is very simple to implement and that leads to a tremendous reduction of the size of the blocks. By using it we gain a huge speed-up for calculations with more than three orbitals. We exploit this speed-up to study the orbital selective Mott transition with SU(2)-symmetric interaction in systems with up to seven orbitals.

The SU(2)-symmetric Kanamori Hamiltonian that is widely used for multi-orbital calculations reads \cite{georgesHundRev}
\begin{align} 
\label{Kanamori}
H_{loc}&=\sum_{a} U n_{a,\uparrow} n_{a,\downarrow} \\
&+\sum_{a>b,\sigma} \Big[U' n_{a,\sigma} n_{b,-\sigma} +  (U'-J) n_{a,\sigma}n_{b,\sigma}\Big]\nonumber\\
&-\sum_{a\ne b}J(d^\dagger_{a,\downarrow}d^\dagger_{b,\uparrow}d^{\phantom{\dagger}}_{b,\downarrow}d^{\phantom{\dagger}}_{a,\uparrow}
+ d^\dagger_{b,\uparrow}d^\dagger_{b,\downarrow}d^{\phantom{\dagger}}_{a,\uparrow}d^{\phantom{\dagger}}_{a,\downarrow} + h.c.) \nonumber.
\end{align}
The index $a$ runs over the $N_{orb}$ orbitals of the impurity, $n_{a,\sigma} \! =\! d^\dagger_{a,\sigma}d^{\phantom{\dagger}}_{a,\sigma}$ is the number operator counting electrons on orbital $a$ with spin $\sigma$. The first term describes the repulsion $U$ for two electrons with opposite spin on the same orbital. In the second line of Eq. (\ref{Kanamori}) one finds the Coulomb interaction $U'$ for two electrons with opposite spin on two different orbitals and $U'-J$, when the spins of the two electrons are aligned. 
The choice $U'=U\!-\!2J$, coming from an exact relation between the parameters for the case of an isolated atom in a cubic crystal-field, is also typically used for realistic calculations.

For our purposes, it is convenient to work in the occupation number basis. For $N_{orb}\!=\!5$, a vector in this basis can be symbolically denoted as follows: 
\begin{equation}
\label{basis}
\begin{array}{cccccccc}
|& \hilite{hiblue}{$\mathbin{\up}\mathbin{\phantom{\dow}}$} &  \hilite{hired}{$\up \dow$} & \hilitenode{hired}{c1}{${\phantom{\up\dow}}$} & \hilite
{hiblue}{$\mathbin{\phantom{\up}}\mathbin{\dow}$} & \hilite{hired}{${\phantom{\up\dow}}$} & \anode{b1}{$\rangle$} & \\
\end{array}
\end{equation}
For reasons that will be clear very soon, we color-code singly-occupied orbitals in blue and empty-/doubly-occupied orbitals in red. 
The ``density-density'' terms that appear in the first two lines of Eq. (\ref{Kanamori}) are diagonal in this basis. 
On the contrary, the two terms contained in the last line of Eq. (\ref{Kanamori}) generate off-diagonal matrix elements.
They are called `spin-flip'' and ``pair-hopping'', respectively, and are needed to preserve the SU(2) spin symmetry: The former flips the spins of two singly-occupied orbitals while the latter transfers a pair of electrons from a doubly-occupied to an empty orbital. 

In the case of many orbitals the size of the basis is pretty big (e.g., for five orbitals $H_{loc}$ is a $1024 \! \times \! 1024$ matrix) but, as we already mentioned, we can reduce $H_{loc}$ to a block diagonal form by using its good quantum numbers.
The most obvious conserved quantities of $H_{loc}$ are the total number of electrons $N$ and the $z$-component of the total spin $S_z$.
In fact, $H_{loc}$ does not connect states with different $N$ and does not change $S_z$ either, since the ``spin-flip'' and the ``pair-hopping'' terms preserve the $z$-component of the total spin.
Also the total spin $\vec{S}^2$ commutes with $H_{loc}$, an obvious consequence of the SU(2) symmetry. Yet, $\vec{S}^2$ turns out not to be practical to implement and therefore is typically not used. The reason for that is the same as for another good quantum number that is not exploited in CT-HYB codes: the ``seniority number''. 
This was introduced by G. Racah \cite{racahPR63} and ``counts'' the number of doubly occupied orbitals in each state. It is easy to see that this is another conserved quantity of $H_{loc}$. However, using $N$, $S_z$ and the ``seniority number'' as quantum labels leads to ambiguities in the definition of the creation and annihilation operators. This can be understood by considering that both $|\! \uparrow, \downarrow, 0 \rangle$ and $| \! \uparrow, 0, \downarrow \rangle$ belong to the same block but $d^\dagger_{2, \uparrow}$ connects them to $| \! \uparrow, \uparrow \downarrow, 0 \rangle$ and $| \! \uparrow, \uparrow,  \downarrow \rangle$ which have different values of the ``seniority number''. 
$\vec{S}^2$ leads to a very similar problem.

Hence, $N$ and $S_z$ are the two quantum numbers typically used in CT-HYB codes.
With this choice, the largest block for, e.g., $N_{orb}\!=\!5$ is $ 100 \! \times \! 100$ and this is still pretty big.
The crucial observation that we make here is the following: 
The Kanamori $H_{loc}$ connects only those states in the occupation number basis that have exactly the same singly-occupied orbitals.
\begin{equation}
\begin{array}{cccccccc}
|& \hilite{hiblue}{$\mathbin{\up}\mathbin{\phantom{\dow}}$} &  \hilite{hired}{$\up \dow$} & \hilitenode{hired}{c1}{${\phantom{\up\dow}}$} & \hilite{hiblue}{$\mathbin{\phantom{\up}}\mathbin{\dow}$} & \hilite{hired}{${\phantom{\up\dow}}$} & \anode{b1}{$\rangle$} & \\
&&&&&&&\\
&&&&&&&\\
|& \hilite{hiblue}{$\mathbin{\up}\mathbin{\phantom{\dow}}$} & \hilite{hired}{${\phantom{\up\dow}}$} & \hilitenode{hired}{c2}{${\phantom{\up\dow}}$} & \hilite{hiblue}{$\mathbin{\phantom{\up}}\mathbin{\dow}$} & \hilite{hired}{$\up \dow$} & \anode{b2}{$\rangle$}&\\
&&&&&&&\\
&&&&&&&\\
|& \hilite{hiblue}{$\mathbin{\phantom{\up}}\mathbin{\dow}$} & \hilite{hired}{${\phantom{\up\dow}}$} & \hilitenode{hired}{c3}{${\phantom{\up\dow}}$} & \hilite{hiblue}{$\mathbin{\up}\mathbin{\phantom{\dow}}$} & \hilite{hired}{$\up \dow$} & \anode{b3}{$\rangle$} & 
\label{Config}
\end{array}
\end{equation}

% draw the arrows between the kets and write the text next to the arrow{c2}s
\tikz[overlay,remember picture] {
  \draw[->,thick,red,dashed,bend left,shorten <= 2pt, shorten >= 2pt] (b1.east) to node[left]{pair-hopping} (b2.east);
  \draw[->,thick,blue,dashed,bend left,shorten <= 2pt, shorten >= 2pt] (b2.east) to  node[left]{spin-flip} (b3.east);
}

Looking at the sketch in Eq. (\ref{Config}) it is clear that neither the ``spin-flip'' nor the ``pair-hopping'' process can turn a singly-occupied orbital into an empty or a doubly-occupied one. This means that the pattern of the singly-occupied orbitals (in other words the list of singly occupied orbitals regardless the spin orientation) is conserved by the Kanamori $H_{loc}$.
Therefore, even though $H_{loc}$ has processes among different orbitals, for each orbital a projector onto single occupations of this orbital (``PS'') commutes with $H_{loc}$. This defines a vector of operators and corresponding quantum numbers
\begin{equation} \label{PS}
{\bf PS} = \left\{ ( n_{a,\uparrow} - n_{a,\downarrow} )^2 \right\} \hspace{0.2cm} \text{for} \hspace{0.2cm} a=1,...N_{orb}. 
\end{equation}
Indeed, $( n_{a,\uparrow} - n_{a,\downarrow} )^2$ yields 0 if the orbital $a$ is either empty or doubly occupied and 1 if the orbital $a$ is singly occupied, proving the projective property. The resulting vector of quantum numbers (for which we employ the same symbol ``PS'') is a binary sequence encoding the information about the pattern of singly occupied orbitals. 
The number (not the pattern) of the singly occupied orbitals has already been previously exploited as good quantum number of Anderson impurity Hamiltonians in Ref. \onlinecite{assaadPRB63}.

\begin{figure}
\begin{overpic}
   [width=\columnwidth]{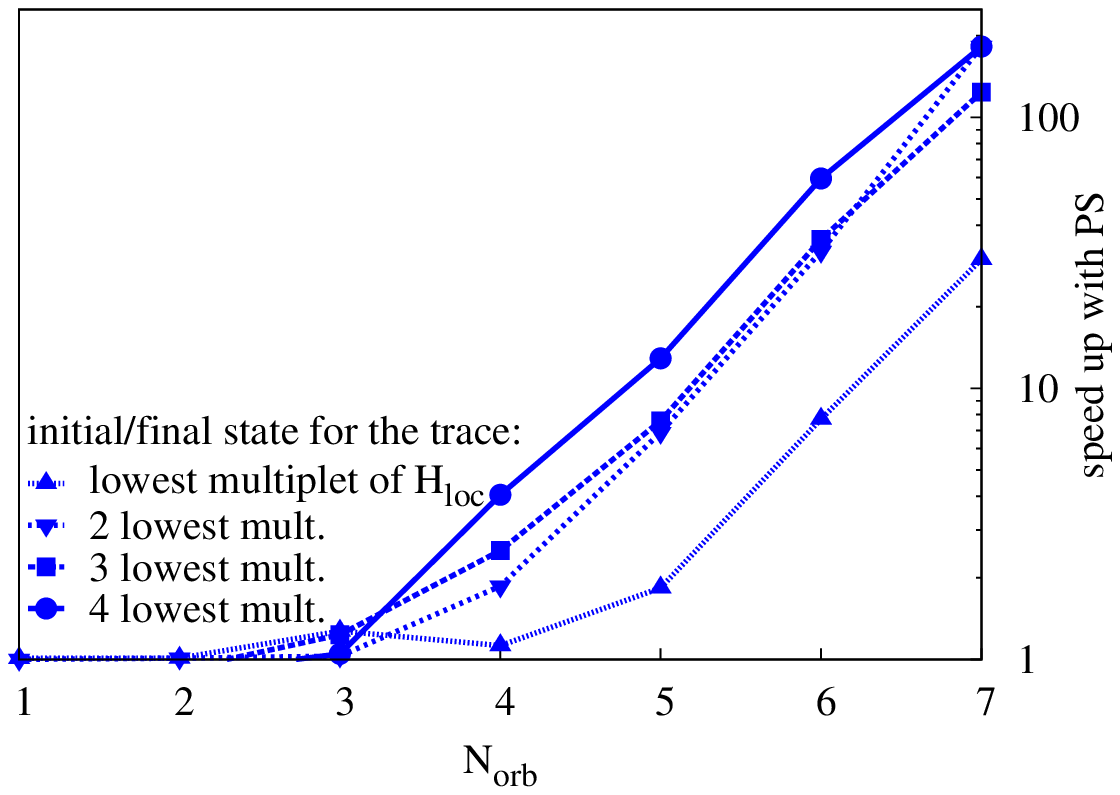}
   \put(-5.15,53.75){
\scalebox{0.85}{
\begin{tabular}{c|c|c|}& $N\;S_z$ & $N\; S_z\; \text{{\bf PS}}$ \\ \hline
$N_{orb}$ & max/mean & max/mean \\
\hline1 & 1/1.00 & 1/1.00\\
2 & 4/1.78 & 2/1.14\\3 & 9/4.00 & 3/1.45\\
4 & 36/10.24 & 6/2.00\\
5 & 100/28.44 & 10/2.90\\
6 & 400/83.59 & 20/4.41\\
7 & 1225/256.00 & 35/6.92\\
\hline
\end{tabular}}}
   \end{overpic}
   \caption{(color online) Table: Maximum and mean block-sizes of the Kanamori Hamiltonian using the total number of electrons in the system $N$, the total spin momentum in $z$-direction $S_z$ (second column) and additionally {\bf PS} (third column) as good quantum numbers, for various $N_{orb}$. Figure: Ratio of CT-HYB runtime (speed up) we obtain for an equal amount of Monte Carlo steps with and without {\bf PS}. Three independent measurement of the runtime were done for each point; the average is plotted and the spread of the results is of the order of the symbols. We first consider the lowest-lying multiplet of $H_{loc}$ as outer state for the trace (good approximation at very low temperatures) and then include progressively more and more of the lowest multiplets. For $N_{orb} \! = \! 7$, the lowest four multiplets cover an energy range of about 4.5 eV, i.e. a range of order $U$.}
   \label{benchmark}
\end{figure}

Labeling blocks of $H_{loc}$ with the set of quantum numbers ($N$,\, $S_z$,\, {\bf PS}) leads to a tremendous reduction of the block size, as shown in the table of Fig. \ref{benchmark}, as well as the size of the $d$ and $d^\dagger$ matrices.
Indeed, any creation and annihilation operator will always connect two blocks in which \emph{all} quantum numbers differ and all the states within the block behave in the same way in this respect.

The most natural implementation of {\bf PS} defined by Eq. (\ref{PS}) is to introduce a single label defined, e.g., in a binary manner as $\sum_a 2^a (n_{a,\uparrow} -n_{a,\downarrow})^2$.
In our code, this information is used to generate $H_{loc}$ in a block diagonal structure.
Already at this stage we can see the improvement gained by using {\bf PS} compared to only $N$ and $S_z$ as conserved quantities. 
In Fig. \ref{benchmark} we show the maximum and the mean block sizes.
The advantage of {\bf PS} becomes striking for system with many orbitals: We obtain block sizes which are in average two orders of magnitude smaller for $N_{orb}\!=\!7$.
We therefore expect a moderate speed up already for four orbitals which should dramatically increase with seven. 
More generally, the number of blocks increases exponentially with the number of orbitals, and the size of the blocks decreases correspondingly.

To demonstrate how much the use of {\bf PS} speeds up actual calculations, we performed single shot simulations on an Anderson impurity model.
This consists of $N_{orb}$ semi-circular bands of half-bandwidth $D\!=\!2$ eV.
The interaction parameters of the Kanamori Hamiltonian were set to intermediate strength values, namely $U\!=\!D$ and $J=0.25U$.
The inverse temperature $\beta$ was set to $100$ eV$^{-1}$ and the chemical potential was set to the half-filling condition $\mu_{HF}=(N_{orb}-\frac{1}{2})U-(N_{orb}-1)\frac{5}{2}J$.
With this model system we performed calculations for $N_{orb}$ varying from 1 to 7 with and without the use of {\bf PS} for otherwise identical parameters as single core jobs on an AMD machine.

As an additional parameter we varied the number of outer eigenstates of $H_{loc}$ over which the fermionic trace is computed. This is a very convenient and clean way of introducing a ``truncation'' parameter in the ``Krylov'' algorithm. It can be understood as follows: At $T\!=\!0$ one can restrict the computation of the trace to the lowest-lying multiplet only. For finite $T$ the calculation is instead exact only upon performing the outer sum over all states of $H_{loc}$, but we observe -- similarly to Ref. \onlinecite{laeuchliPRB80} -- that the calculation converges rapidly upon including more and more of the lowest-lying multiplets of $H_{loc}$. 

In Fig. \ref{benchmark} we show the ratio of the QMC runtime with and without {\bf PS}. This demonstrates that, as expected, the advantage of using {\bf PS} is huge for calculations with large number of orbitals. In the Figure, the average of three independent timings for each value of $N_{orb}$ are plotted. 
If we look at the curves in which the initial and final states for the trace are not restricted to the lowest-lying multiplet (i.e. the typical situation for calculations at room temperature) we obtain a performance gain of one order of magnitude for five and a really remarkable gain of two orders of magnitude for seven orbitals. 
This makes self-consistent DMFT calculations for such systems really accessible.
It also enables us to check the convergence of DMFT calculations for multi-orbital systems with respect to the number of multiplets as outer states for the trace, which was previously not always possible since the simulation was too costly. 
In addition, this also allows us to explore parameter regions which were formerly prohibitively expensive.

\begin{figure}[t]
   \begin{tikzpicture}
   \node[anchor=south west,inner sep=0] (image) at (0,0) {\includegraphics[width=\columnwidth]{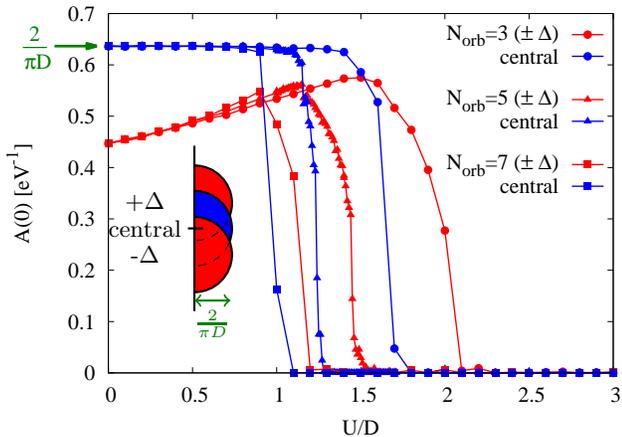}};
    \begin{scope}[x={(image.south east)},y={(image.north west)}]
      \node(cb) at (0.22,0.47) [minimum width=1.3cm,]{central};
      \node(rcb)[minimum width=1.3cm,inner sep=0cm,right = 0.25cm of cb]{};
      \node(acb)[minimum width=1.3cm,inner sep=0cm,above = 0.25cm of cb]{};
      \node(bcb)[minimum width=1.3cm,inner sep=0cm,below = 0.25cm of cb]{};
      \node(ub) at ($ (cb) + (0,0.35cm) $)[minimum width=1.3cm]{+$\Delta$};
      \node(rub)[minimum width=1.3cm,inner sep=0cm,right = 0.25cm of ub]{};
      \node(aub)[minimum width=1.3cm,inner sep=0cm,above = 0.25cm of ub]{};
      \node(bub)[minimum width=1.3cm,inner sep=0cm,below = 0.25cm of ub]{};
      \node(lb) at ($ (cb) - (0,0.35cm) $)[minimum width=1.3cm,]{-$\Delta$};
      \node(rlb)[minimum width=1.3cm,inner sep=0cm,right = 0.25cm of lb]{};
      \node(alb)[minimum width=1.3cm,inner sep=0cm,above = 0.25cm of lb]{};
      \node(blb)[minimum width=1.3cm,inner sep=0cm,below = 0.25cm of lb]{};
      \node(ini) at ($ (cb) + (0,1.1cm) $)[minimum width=1.3cm]{};
      \node(fin) at ($ (cb) - (0,1.1cm) $)[minimum width=1.3cm]{};
      \node(fin1)[minimum width=1.3cm,inner sep=0cm,above = 0.0005cm of fin]{};
      \node(fin2)[right = 0.25cm of fin1]{};
      \draw[thick](ini.east) -- (fin.east);
      \draw[fill=red,thick] (aub.east) -- (bub.east) arc (-90:90:0.5cm);
      \draw[fill=blue,thick] (acb.east) -- (bcb.east) arc (-90:90:0.5cm);
      \draw[fill=red,thick] (alb.east) -- (blb.east) arc (-90:90:0.5cm);
      \draw[dashed] (aub.east) -- (bub.east) arc (-90:90:0.5cm);
      \draw[dashed] (acb.east) -- (bcb.east) arc (-90:90:0.5cm);
      \draw[<->,darkgreen,thick] (fin1.east) to node[below]{$\frac{2}{\pi D}$} (fin2.east); 
      \node(tick)[draw,thick,minimum width=0.2cm,inner sep=0cm,minimum height=0cm,right = -0.1cm of cb]{};
    \end{scope}
   \end{tikzpicture}
  \caption{(color online) Spectral weight at the Fermi level $A(0)$ for the model sketched in the inset with one symmetric band (blue) and the remaining ones symmetrically shifted above and below of $\Delta \! = \! 0.7D$. This model as been solved for $N_{orb} \! = \! 3$ in Ref. \onlinecite{deMediciPRB83}. We clearly observe an orbital selective region, with the central band Mott insulating (zero spectral weight) and the shifted bands still metallic (nonzero spectral weight) in a finite $U$-interval up to $N_{orb} \! = \! 7$.}
\label{transition}
\end{figure}

To demonstrate the practical advantages of using {\bf PS} we apply our implementation of the CT-HYB to a three-, five- and seven-orbital model system, as the one sketched in the bottom-left corner of Fig. \ref{transition}. 
This model is ideal for studying the interaction-driven orbital selective Mott transition (OSMT), as shown in Ref. \onlinecite{deMediciPRB83}.
It differs from the more commonly used model with bands of different widths, since it consists of one central orbital associated to a symmetric band and one, two or three orbitals shifted up in energy by $\Delta\!=\!0.7D$, where $D$ is half the bandwidth, and an equal number of orbitals shifted to lower energies by the same $\Delta$.
All calculations were performed with the interaction (\ref{Kanamori}), at half-filling, with $J\!=\!0.25U$ and $\beta D\!=\!100$.
The DMFT self-consistency was reached considering only the lowest-lying multiplet as outer states in the trace. The stability of the solution versus the inclusion of more multiplets was afterwards checked. 

Hitherto, model studies of the orbital selective Mott transition with DMFT have focused almost exclusively on $N_{orb} \! \leq\! 3$. Here we want to test the robustness of the OSMT against the number of orbitals. For that we compare the cases of $N_{orb} \! =\! 3$, $5$ and $7$. The only two calculations with five orbitals we are aware of are the ones of Refs. \onlinecite{deMediciJSNM22} and of Ref. \onlinecite{laeuchliPRB80}. Both were done for somewhat different models than the one considered here but, more importantly, the former was carried out with a simplified slave-spin mean-field solver while the latter addressed the filling-driven OSMT only. 

Our findings are summarized by the data shown in Fig. \ref{transition}. For $N_{orb} \! =\! 3$ we reproduce the transition values reported in Ref. \onlinecite{deMediciPRB83} and we find the existence of a similar, though somewhat smaller, orbital selective region for $N_{orb} \! =\! 5$ and 7. 
We can therefore conclude that in a model with SU(2)-invariant interaction characterized by one symmetric band and four or six other ones symmetrically shifted in energy an orbital selective region exists in which the central band gets insulating (its spectral weight at the Fermi level $A(0)$ vanishes), while the shifted bands stay metallic (finite $A(0)$).

In Fig. \ref{transition} one can see that the critical $U$ dividing the metallic and the orbital selective regions decreases with $N_{orb}$. This is a consequence of the effect of the sizable value for the Hund coupling $J$ used. For $J \! = \! 0$ we would have observed the opposite because more orbitals lead to a larger mobility and therefore a larger critical value for the transition to the insulating state. This delocalizing effect is counteracted by the presence of a large Hund coupling which strongly suppresses orbital fluctuations increasing the insulating region. This is in agreement with what was reported in Refs. \onlinecite{deMediciPRB83,hauleNJP11,shorikov2008}

In conclusion, we propose invariants for SU(2)-symmetric Kanamori Hamiltonians, i.e. the single occupation of each orbital. We introduce a related quantum label leading to very small blocks of the matrices. This results in a speed-up of CT-HYB quantum Monte Carlo calculations of up to two orders of magnitude, and allows us to study much more comfortably models with a large number of orbitals. 
As an example we have considered the interaction-driven orbital selective Mott transition at half-filling and found that it persists up to seven orbitals.
In addition to the class of problems for which {\bf PS} is useful, there are cases in which more complete schematization of the full Coulomb repulsion are needed. In particular the richer multiplet structure of ``Slater''-type of parametrizations of the Coulomb interaction can play a role in some realistic DMFT calculations with five or more orbitals. 
In order to flexibly study such very complex Hamiltonians with CT-HYB, good quantum numbers as effective as {\bf PS} would be immensely helpful.

\emph{Acknowledgments} -- N.P. and G.S. are indebted to M. Ferrero, E. Gull and P. Werner for help and feedback in writing the SU(2)-symmetric code. We also thank L. de'~Medici and G. Rohringer for fruitful discussions and F. Assaad for drawing our attention to Ref. \onlinecite{assaadPRB63} while writing this manuscript. M. Capone kindly told us about some similar (unpublished) ideas used for the calculations of Ref. \onlinecite{caponePRL86}. G.S. and A.T. would also like to acknowledge the hospitality of the people of Campello sul Clitunno and its inspiring atmosphere. This work has been supported in part by the Research Unit FOR 1346 of the DFG (FWF Project ID I597-N16).

\end{document}